\documentclass[aps,pre,reprint,superscriptaddress,showkeys]{revtex4-1}

\usepackage[ngerman, english]{babel} 
\usepackage[utf8]{inputenc} 
\usepackage[T1]{fontenc} 

\usepackage{xcolor}
\usepackage{epstopdf}
\usepackage{graphicx}
\usepackage{csquotes}
\usepackage{tikz}

\usepackage{amsmath}
\usepackage{amssymb}
\usepackage{amsfonts}
\usepackage{amsthm}
\usepackage{bm}
\usepackage{mathdots}

\usepackage[hidelinks]{hyperref} 
\usepackage{calc} 
\usepackage{enumitem} 
\usepackage[caption=false]{subfig}
\usepackage{glossaries}
\newacronym{slelgc}{SLE-LGC}{Stuart-Landau ensemble with linear global coupling}
\newacronym{slenlgc}{SLE-NLGC}{Stuart-Landau ensemble with nonlinear global coupling}

\newcommand{\tum}{Physik-Department, Nonequilibrium Chemical Physics,
	Technische Universität München, D-85748 Garching, Germany
}

\begin{document}


\title{Lyapunov Analysis of Chimera States in Globally Coupled Stuart-Landau oscillators}


\author{Kevin H\"ohlein}
\affiliation{\tum}
\author{Felix P. Kemeth}
\affiliation{\tum}
\author{Katharina Krischer}
\affiliation{\tum}


\date{\today}

\begin{abstract}
Oscillatory systems with long-range or global coupling offer promising insight into the interplay between high-dimensional (or microscopic) chaotic motion and collective interaction patterns. Within this paper, we use Lyapunov analysis to investigate whether chimera states in globally coupled Stuart-Landau (SL) oscillators exhibit collective degrees of freedom. We compare two types of chimera states, which emerge in SL ensembles with linear and nonlinear global coupling, respectively, the latter introducing a constraint that conserves the oscillation of the mean. Lyapunov spectra reveal that for both chimera states the Lyapunov exponents split into different groups with different convergence properties in the limit of large system size. Furthermore, in both cases the Lyapunov dimension is found to scale extensively and the localization properties of covariant Lypunov vectors manifest the presence of collective Lyapunov modes. Here, however, we find qualitative differences between the two types of chimera states: Whereas the ones in the system under nonlinear global coupling exhibit only slow collective modes corresponding to Lyapunov exponents equal or close to zero, those which experience the linear mean-field coupling exhibit also faster collective modes associated with Lyapunov exponents with large positive or negative values.
\end{abstract}


\maketitle

\section{Introduction\label{sec:intro}}

 Chaos readily emerges in systems composed of many coupled oscillatory 'units'. These units may represent individual oscillators or infinitesimally small patches of spatially extended oscillatory media, and the coupling is what gives rise to chaotic instabilities. The origin of the instabilities, however, might lie in microscopic,  i.e. local, interactions, involving only a small number of oscillators, or may result from complex interaction patterns on macroscopic system scales. The character of the corresponding chaotic dynamics, accordingly, may vary significantly.
 
 
 One way to shine light on the nature of the chaotic dynamics is to study chaotic states in oscillatory ensembles containing different numbers of oscillators $N$ while being subject to equivalent coupling schemes. Thereby, the chaotic dynamics are usually characterized in terms of Lyapunov exponents (LEs), which measure the average infinitesimal divergence rates of the motion in phase space~\cite{Oseledets1968,Pikovsky2016}. Our understanding of chaos within this context is comparatively advanced in two limiting cases, so-called intensive and solely extensive chaos. Intensive chaos is characterized by an (in general small) number of positive LEs that is independent of the number of oscillatory units. It occurs, e.g., for certain parameters in systems of globally coupled oscillators ~\cite{Matthews1991,Hakim1992,Nakagawa1993}. 
 In the opposite case, when the chaotic motion arises exclusively from microscopic degrees of freedom and the number of positive LEs scales linearly with $N$, then the chaotic dynamics is solely extensive. Such behavior was shown to exist in some generic spatially one-dimensional models~\cite{Livi1986}.  
 However, it has been recently established that large systems with global couplings schemes or spatially extended systems in two or three spatial dimensions, might exhibit chaotic motion that belongs to neither of these two situations. 
 Instead, their dynamics is characterized by both a few, collective macroscopic modes and a large number of microscopic, chaotic degrees of freedom~\cite{Matthews1991,Hakim1992,Nakagawa1993}. 
 Our knowledge how and when such collective modes develop and which role they play in the overall high-dimensional chaotic dynamics is still very limited.
 
The investigation of collective dynamics in high-dimensional chaotic systems remains a challenge. 
For a long time, it was common belief that conventional Lyapunov analysis cannot capture these collectively chaotic dynamics correctly~\cite{Shibata1998}.
Instead, finite-size perturbations needed to be studied in order to identify collective dynamical modes and the associated LEs. 
Only recently, Takeuchi et al.~\cite{Takeuchi2009,Takeuchi2013} were able to demonstrate that standard Lyapunov analysis can in fact provide information on collective dynamics. The key of their analysis was the calculation of the covariant Lyapunov vectors (CLVs) associated to the LEs~\cite{Oseledets1968,Ginelli2007,Wolfe2007}. 
Having access to CLVs, Takeuchi et al.\ complemented the information provided by the Lyapunov spectrum with an investigation of the localization or delocalization properties of CLVs associated with particular LEs. 
They illustrated their strategy for $N$ globally coupled Stuart-Landau (SL) oscillators in a high-dimensional chaotic state. 
Most of the CLVs were found to be well localized for a large range of ensemble sizes $N$, while a small number of modes appeared to become increasingly delocalized with increasing $N$. 
The respective perturbations, named collective Lyapunov modes, were shown to be related to macroscopically chaotic degrees of freedom. 
Moreover, in accordance with earlier work on collective chaos by Nakagawa and Kuramoto \cite{Nakagawa1993,Nakagawa1994,Nakagawa1995}, they showed that the overall Lyapunov spectrum can be separated into parts with different convergence properties in the limit of large $N$. 
In particular, they identified an extensively scaling group of $\mathcal{O}\!\left(N\right)$ positive LEs in the middle of the spectrum, resulting in a flattening of the spectrum. 
Around the most positive and most negative exponents, however, they also observed sub-extensive groups of exponents, the number of which was shown to scale approximately as $\mathcal{O}\!\left(\ln\!\left(N\right)\right)$~\cite{Takeuchi2013}. \\

In this paper, we perform a related Lyapunov analysis of chimera states in globally coupled SL oscillators. Being composed of coexisting groups exhibiting synchronous and asynchronous motion, respectively, chimera states are peculiar dynamic states which can be seen as 'a natural link between coherence and incoherence'~\cite{Omelchenko2008}. 
They might offer insight into natural phenomena such as some neural activity patterns~\cite{Sakaguchi2006} or hydrodynamic flows with mixed laminar and turbulent patterns~\cite{Barkley2005}. 
So far, there are only a few studies on Lyapunov analysis of chimera states that mainly focus on systems of coupled phase oscillators.
Wolfrum et al.~\cite{Wolfrum2011,Wolfrum2011b} showed that the Lyapunov spectra for chimera states in coupled phase oscillator networks of finite size exhibit an extensive number of positive LEs, revealing the hyperchaotic nature of chimera states. 
Yet, all the positive exponents decay with increasing system size and finally yield neutrally stable zero exponents in the limit $N\to\infty$. 
Nevertheless, they found the Lyapunov dimension $D_\text{L}$ to scale extensively with system size. 
Furthermore, Botha et al.~\cite{Botha2016,Botha2018} studied the distribution of finite-time LEs in chimera-like dynamics and identified characteristic patterns in their temporal distribution function. 
While for usual chaotic states without synchronization pattern the finite-time LEs follow a Gaussian distribution, they found that the distribution in the case of chimera states possesses a complex, multi-modal shape, which they proposed to use as an indicator for chimera-like behavior.\\

In this work, however, we want to focus our attention on properties of the asymptotic Lyapunov spectra, without further reference to the temporal distribution of finite-time exponents. In particular, we examine whether Lyapunov spectra for chimera states of the globally coupled SL ensemble exhibit similar patterns as observed for the above-mentioned chaotic states, which exist in the same type of oscillatory network for different parameter values~\cite{Takeuchi2009}.  
Furthermore, we compare chimera states in these mean-field coupled SL ensembles with those arising in SL oscillators subject to a nonlinear global coupling that conserves a harmonic oscillation of the ensemble mean. This coupling scheme was introduced to describe experiments on Si electrodissolution~\cite{Schmidt2014,Schoenleber2014}. 
In our context, it gives additional information of how a global constraint might further impact the interplay between high-dimensional incoherent motion and collective interaction patterns.\\

The paper is organized as follows. In the next section, we introduce our two models as well as the particular types of chimera states that are subject to our investigations. 
In section \ref{sec:methods}, the notation used throughout the paper is introduced, Lyapunov analysis as performed on our data reviewed, and measures used to characterize the dynamics, in particular the Lyapunov dimension $D_\text{L}$ and the inverse participation ratio (IPR) are defined. 
In the results and discussion section \ref{sec:results}, we start by investigating the Lyapunov spectra of the two types of chimera states for $N=16$ oscillators in detail, discuss then how the spectra change when $N$ is stepwise doubled up to $N=256$, and end with a discussion of how the Lyapunov dimension and the IPR depend on $N$. The latter allows us in particular to draw conclusion on the existence and on some properties of collective Lyapunov modes. The paper is completed with the conclusion section \ref{sec:discussion}.

\section{Dynamical systems\label{sec:systems}}

\begin{figure}
	\includegraphics[]{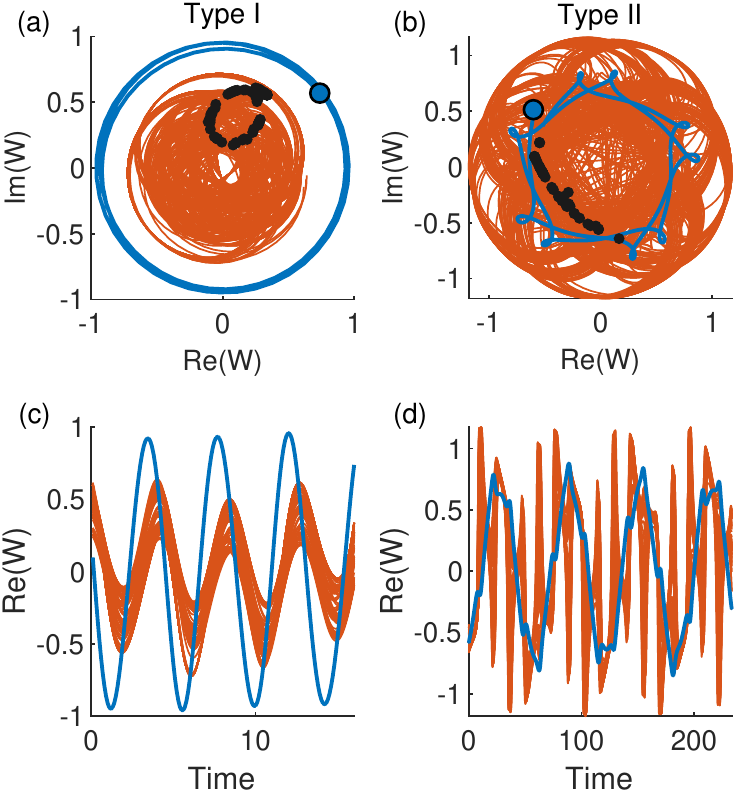}
	\caption{Simulation data of the type I chimera, (a) and (c), and the type II chimera state, (b) and (d), with $N=256$ oscillators.
          Trajectories of coherent and incoherent oscillators are shown in blue and red, respectively.
          Top: Oscillator trajectories in the complex plane. Dots represent a snapshot of the oscillator states. The coherent group of oscillators is shown in blue, whereas incoherent oscillators are colored black. Bottom: Real part of the oscillator states as a function of time.}
\label{fig:dynamics}        
\end{figure}

There is a vast number of different chimera states that have been reported in literature in recent years~\cite{Panaggio2015, Kemeth2016, Scholl2016, Omelchenko2018}.
In this article, we focus on two kinds, also called type I and type II chimeras~\cite{Schmidt2015}, which can be observed in systems with long-range interactions~\cite{Sethia2014, Schmidt2014}.
The former, the type I chimera, appears in Stuart-Landau ensembles with linear global coupling. The time evolution of the systems is governed by equations of the form
\begin{equation}
  \partial_t W_k = W_k - \left(1+ic_2\right)\left|W_k\right|^2 W_k + \kappa \left(\frac{1}{N} \sum_{j=1}^N W_j - W_k\right),
  \label{eq:slelgc}
\end{equation}
wherein $W_k$ denotes the complex amplitude of oscillator $k$, and $k=1,\dots,N$, with $N$ being the total number of oscillators.
Parameters are the real shear, $c_2$, and the complex coupling constant, $\kappa $, which we set to $c_2=2$ and $\kappa = 0.7\,(1-i)$ in the following. Hereby,  $i$ denotes the imaginary unit. Given this set of parameters, we investigate ensemble sizes $N=2^l$ for $4\le l \le 8$, and find the type I chimera state to be a stable attractor for $N\ge16$. The oscillator dynamics in the complex plane, as well as the corresponding time series of the real parts are shown in Fig.~\ref{fig:dynamics}~(a) and~(c), respectively, for $N=256$. 
There, one observes a coherent group of oscillators, depicted in blue, with a larger absolute value of $W_k$, and an incoherent group composed of oscillators with smaller amplitudes, shown in red. The thick dots highlight a snapshot of the dynamics.  Note that the motion of the incoherent group, when viewed in the complex plane, lies on a string-like object undergoing stretching and folding. The structure is similar to a Birkhoff-Shaw attractor~\cite{Thompson1984}, and resembles the chaotic motion found in mean-field coupled SL ensembles~\cite{Nakagawa1994,Takeuchi2009}.\\

In contrast to the type I chimeras, the type II chimera state has only been observed in systems with nonlinear global coupling~\cite{Schmidt2015}.
In particular, it appears in Stuart-Landau ensembles of the form
\begin{align}
  \partial_t W_k & = -i\nu W_k  - \left(1+i\nu\right)\left(\frac{1}{N} \sum_{j=1}^N W_j - W_k\right) \nonumber\\
  & - \left(1+ic_2\right)\left(\frac{1}{N} \sum_{j=1}^N\left|W_j\right|^2 W_j-\left|W_k\right|^2 W_k\right),
  \label{eq:slenlgc}
\end{align}
with parameters $c_2 = -0.66$, $\nu = 0.1$ and the initial absolute value of the mean amplitude, 
\begin{equation*}
	\left\lvert\left\langle W \right\rangle\right\rvert = \bigg\lvert\sum_jW_k(t=0)/N\bigg\rvert=\eta = 0.67\text{.}
\end{equation*}
Exemplary simulation data are depicted in Fig.~\ref{fig:dynamics}~(b) and~(d) for $N=256$.
The coloring scheme is identical to that of the type I chimera state.
Note that the type II chimera state exhibits an additional frequency component in the oscillator dynamics and that there is no clear amplitude separation between the coherent and incoherent oscillators, as opposed to the type I dynamics.
Furthermore, qualitative differences between both types of chimera states can be captured in terms of order parameters~\cite{Kemeth2016}: While the type II chimera states possess an order parameter with oscillatory behavior, leading to categorization as 'breathing chimeras', the order parameter of the type I chimera state fluctuates irregularly around a constant value, as characteristic for the class of 'turbulent chimeras'.
In the following, we investigate the ensemble dynamics of the two cases more carefully using Lyapunov analysis, with a particular emphasis on the apparently chaotic motion of the oscillators in the incoherent groups.

\newcommand{\xref}{\hat{\bm{x}}}
\newcommand{\wref}{\hat{\bm{W}}}
\newcommand{\R}{\mathbb{R}}
\newcommand{\N}{\mathbb{N}}
\section{Background and methods\label{sec:methods}}

\subsection{Lyapunov analysis\label{sec:lyapunov}}

Lyapunov analysis provides quantitative methods to investigate the degree of chaoticity of dynamical systems \cite{Pikovsky2016}. To introduce concepts and notation, we consider a sufficiently well-behaved dynamical system, whose time evolution is described by a time-dependent vector $\bm{x}=\bm{x}\!\left(t\right)\in\R^{2N}$, and governed by the autonomous ordinary differential equation
\begin{equation}
	\dot{\bm{x}} = \bm{f}\!\left(\bm{x}\right)\text{.}
	\label{eq:ode}
\end{equation}
Eqs.~\eqref{eq:slelgc} and~\eqref{eq:slenlgc} can be cast in this form by introducing the real-valued oscillator coordinates $\left\{a_k,b_k\right\}_{k=1}^N$ through  
\begin{align*}
	a_k = \frac{1}{\sqrt{2}}(W_k + W_k^*)\text{,}\\
	b_k = \frac{1}{\sqrt{2}}(W_k - W_k^*)\text{,}
\end{align*}
wherein $W_k^*$ denotes the complex conjugate of $W_k$. Organizing all coordinates in a $2N$-dimensional state vector then yields
\begin{equation*}
	\bm{x} = \left(\begin{array}{c}
	\bm{a} \\ \bm{b}
	\end{array}\right) = \left(a_1,\ldots,a_N,b_1,\ldots,b_N\right)^T\text{.}
\end{equation*}
Furthermore, we specify a reference trajectory $\xref\!\left(t\right)$, which at any point in time has to satisfy Eq.~\eqref{eq:ode}, and investigate the growth or decay of small perturbations around $\xref\!\left(t\right)$. The time evolution of infinitesimal perturbations $\delta\bm{x}=\delta \bm{x}\!\left(t\right)\in T_{\xref\left(t\right)}\R^{2N}$ is governed by the tangent-space dynamics
\begin{equation}
\delta\dot{\bm{x}} = \left.\frac{\partial\bm{f}}{\partial\bm{x}}\right\rvert_{\xref\left(t\right)}\delta \bm{x}\text{,}\label{eq:tanspace}
\end{equation}
wherein $\left.\frac{\partial\bm{f}}{\partial\bm{x}}\right\rvert_{\xref\left(t\right)}$ is the Jacobian matrix of $\bm{f}\!\left(\bm{x}\right)$, evaluated at $\xref\!\left(t\right)$, and $T_{\xref\left(t\right)}\R^{2N}$ denotes the tangent space of the dynamical system in the vicinity of $\xref\!\left(t\right)$. For our purposes, we have $T_{\xref\left(t\right)}\R^{2N}\cong\R^{2N}$. 

Given a suitable initial condition $\delta\bm{x}\!\left(t_0\right) = \delta\bm{x}_0 \in \R^{2N}$, Eq.~\eqref{eq:tanspace} possesses the analytic solution
\begin{equation}
\delta \bm{x}\!\left(t\right) = \Psi_{\xref}\!\left(t,t_0\right)\delta \bm{x}_0\text{,}
\label{eq:soltanspace}
\end{equation}
wherein $ \Psi_{\xref}\!\left(t,t_0\right)$ denotes the fundamental matrix of Eq.~\eqref{eq:tanspace}, encoding the time evolution of perturbation vectors from $T_{\xref\left(t_0\right)}\R^{2N}$ at time $t_0$ to $T_{\xref\left(t\right)}\R^{2N}$ at time $t$. Using Eq.~\eqref{eq:soltanspace}, as well as the standard scalar product in $\R^m$, the squared amplitude of some perturbation vector $\delta\bm{x}\!\left(t\right)$ can thus be expressed as 
\begin{equation*}
	\left\lVert\delta\bm{x}\!\left(t\right)\right\rVert^2 = \delta\bm{x}_0^T\,\Psi_{\xref}\!\left(t,t_0\right)^T\!\Psi_{\xref}\!\left(t,t_0\right)\delta \bm{x}_0\text{.}
\end{equation*}
Oseledets' theorem,  also known as the multiplicative ergodic theorem \cite{Oseledets1968}, states that under rather general conditions there exist symmetric matrices
\begin{equation}
	M_{\xref\left(t_0\right)}^{(\pm)} = \lim_{t\to \pm\infty} \left[\Psi_{\xref}\!\left(t,t_0\right)^T\!\Psi_{\xref}\!\left(t,t_0\right)\right]^{\frac{1}{2 t}}\text{,}
	\label{eq:lyapmatrix}
\end{equation}
with real positive eigenvalues $\mu_1>\mu_2>\ldots>\mu_r$, $r\le2N$, which thus characterize the long-term expansion and contraction rates of perturbation vectors. Note here that symmetries of the system as well as of the dynamical pattern can give rise to degeneracies in the spectrum, which we will discuss later on in more detail. Denoting with $g_k$ the degree of degeneracy of eigenvalue $\mu_k$, the degeneracies satisfy $\sum_{k=1}^rg_k=2N$. Thereby, the degeneracies $g_k$ also reflect the dimensionality of the eigenspaces $U_{\xref\left(t_0\right),k}^{(\pm)}$ of $M_{\xref\left(t_0\right)}^{(\pm)}$, associated with eigenvalue $\mu_k$. Furthermore, the eigenvalues $\mu_k$ and degrees of degeneracy $g_k$ coincide for $M_{\xref\left(t_0\right)}^{(+)}$ and $M_{\xref\left(t_0\right)}^{(-)}$, and are independent of the particular reference trajectory for ergodic dynamics. The Lyapunov exponents $\lambda_1>\lambda_2>\ldots>\lambda_r$ are then defined through $\lambda_k = \ln \mu_k$ and satisfy
\begin{equation}
	\left\lVert \Psi_{\xref}\!\left(t,t_0\right) \delta \bm{x}_0\right\rVert \sim e^{\lambda_kt}\left\lVert\delta \bm{x}_0\right\rVert\text{,}
	\label{eq:asstanspace}
\end{equation}
for $\delta \bm{x} \in \Omega_{\xref\left(t_0\right),l}$. Therein, the subspaces $\Omega_{\xref\left(t_0\right)}^{(k)}\subset T_{\xref\left(t_0\right)}\R^{2N}$ are obtained from suitable intersections of eigenspaces of $M_{\xref\left(t_0\right)}^{(+)}$ and $M_{\xref\left(t_0\right)}^{(-)}$. Employing the  above notation, we define
\begin{align*}
	\Gamma_{\xref\left(t_0\right),k}^{(+)} &= \bigoplus_{l=k}^r \, U_{\xref\left(t_0\right),l}^{(+)}\text{,}\\
	\Gamma_{\xref\left(t_0\right),k}^{(-)} &= \bigoplus_{l=1}^k \, U_{\xref\left(t_0\right),l}^{(-)}\text{,}
\end{align*}
with  $\oplus$ denoting the direct sum of vector spaces, and obtain
\begin{equation*}
	\Omega_{\xref\left(t_0\right),k} = \Gamma_{\xref\left(t_0\right),k}^{(+)}\cap\Gamma_{\xref\left(t_0\right),k}^{(-)}\text{.}
\end{equation*}
The set of subspaces $\left\{\Omega_{\xref\left(t_0\right),k}: 1\le k \le r \right\}$ is called Oseledets' splitting \cite{Oseledets1968, Ginelli2007} and provides a non-orthogonal decomposition of the tangent space according to different expansion rates of infinitesimal perturbations, i.~e.\
\begin{equation*}
	T_{\xref\left(t_0\right)}\R^{2N} = \bigoplus_{k=1}^r \, \Omega_{\xref\left(t_0\right),k}\text{.}
\end{equation*}
 The Oseledet subspaces satisfy $\dim\left(\Omega_{\xref\left(t_0\right),k}\right) = g_k$. In contrast to the eigenspaces $U_{\xref\left(t_0\right),l}^{(\pm)}$ of $	M_{\xref\left(t_0\right)}^{(\pm)}$, Oseledets' splitting is norm-independent, invariant under time inversion, and depends on the current system state in a way that is covariant with respect to the dynamical flow, i.~e.
\begin{equation*}
\Omega_{\xref\left(t\right),k} = \Psi_{\xref}\!\left(t,t_0\right) \Omega_{\xref\left(t_0\right),k}\text{.}
\end{equation*}
The spanning vectors of Oseledets' splitting are called covariant Lyapunov vectors (CLVs)~\cite{Ginelli2007,Kuptsov2012,Ginelli2013}, and by virtue of Eq.~\eqref{eq:asstanspace} indicate the local orientation of stable and unstable manifolds in phase space.
The spanning vectors of the orthogonal eigenspaces of $M_{\xref\left(t_0\right)}^{(\pm)}$ are known as forward (FOLVs, $(+)$) and backward orthogonal Lyapunov vectors (BOLVs, $(-)$). Despite the superior dynamical properties of CLVs compared to FOLVs and BOLVs, the efficient computation of CLVs has become possible only recently due to algorithms by Ginelli et al.~\cite{Ginelli2007}, Wolfe and Samelson~\cite{Wolfe2007}, and later on Kuptsov and Parlitz~\cite{Kuptsov2012}, who improved on the method by Wolfe and Samelson. CLVs have since attracted a large amount of scientific interest and proved to be a fruitful source of insight, especially into phase-space structures of high-dimensional dynamical systems. In particular, CLVs have been used to study e.~g.\ dynamics of rigid disk systems \cite{Bosetti2010,Truant2014}, chaotic motion in spatially extended systems \cite{Yang2013,Xu2016}, stability properties of geophysical models~\cite{Vannitsem2016} and collective chaos in systems of coupled oscillators \cite{Takeuchi2009,Takeuchi2011,Takeuchi2013}.

\subsection{Lyapunov dimension and inverse participation ratio}

The dimensionality of a chaotic attractor can be estimated from the Lyapunov dimension
\begin{equation}
D_\text{L} := L+ \frac{\sum_{l=1}^{L}\lambda_l}{\left\lvert\lambda_{L+1}\right\rvert}\text{.}
\label{eq:lyapdim}
\end{equation}
with $L\le 2N$ denoting the largest integer for which $\sum_{l=1}^{L}\lambda_l>0$. Note here that for computing $D_\text{L}$, we have to take account for degeneracies in the Lyapunov spectrum explicitly. The summation in Eq.~\eqref{eq:lyapdim} therefore runs over an extended index $l\in\left\{1,\ldots,2N\right\}$, which enumerates all of the $2N$ potentially degenerate LEs in a way that satisfies $\lambda_{1}\ge\lambda_2\ge\ldots\ge\lambda_{2N}$. Under generic circumstances, the Kaplan-Yorke conjecture then states that $D_\text{L}$ is an upper bound of the information dimension $D_\text{I}$ of the dynamical pattern~\cite{Kaplan1979}, which is obtained as a special case of the Renyi-$q$ dimension for $q=1$~\cite{Renyi1970,Grassberger1983}, and is closely related to the information production in the underlying system~\cite{Ott1997}. A linear scaling of $D_\text{L}$ with the system size, i.~e.\ in our case the number of oscillators $N$, is commonly used to demonstrate extensivity of chaotic dynamics.

As mentioned above, oscillatory systems with global coupling schemes might also possess collective modes arising from strong correlations~\cite{Matthews1991,Hakim1992,Nakagawa1993}, which do not scale linearly with system size. Such collective modes can be identified investigating the localization or delocalization properties of CLVs associated with particular LEs.  
Using oscillator coordinates and the above notation, an arbitrary perturbation can be expressed as
\begin{equation}
\delta \bm{x} = \left(\delta a_1,\ldots, \delta a_N, \delta b_1, \ldots, \delta b_N \right)^T\in\R^{2N}\text{,}
\label{eq:pert}
\end{equation}
wherein $\delta a_k$ and $\delta b_k$ denote the relative perturbation amplitudes affecting the real and imaginary part of the oscillator state $W_k$, respectively. As demonstrated by Takeuchi et al.~\cite{Takeuchi2009}, the IPR defined by Mirlin et al.~\cite{Mirlin2000} is a suitable measure for vector localization. For perturbations in the form of Eq.~\eqref{eq:pert}, the IPR can be written as
\begin{equation}
\text{IPR} =  \sum_{k = 1}^{N}\left(\delta a_k^2+\delta b_k^2\right)^2\text{.}
\label{eq:ipr}
\end{equation}
Therein, we assume vector normalization according to $\sum_{k = 1}^N\left(\delta a_k^2 + \delta b_k^2\right) = 1$. By definition, the IPR then takes values between  $N^{-1}$ and $1$. Large values are obtained if the vector under study possesses a small number of large-amplitude components, indicating localization of the perturbation mode. Smaller values are obtained if all terms in the summation are of similar magnitude so that a greater number of oscillators is affected by the perturbation. A small IPR value is thus indicative of a collective Lyapunov mode.

\subsection{Numerical methods and simulation details}

For the integration of the system dynamics, we use the variable-step Dormand-Prince method \cite{Dormand1980} implemented in the explicit Matlab integration function ode45 \cite{Shampine1997} with a maximum time step of $5\cdot10^{-3}$ for Eq.~\eqref{eq:slelgc} and $1.5\cdot 10^{-2}$ for Eq.~\eqref{eq:slenlgc}. During simulation of the type I chimera states, we allowed an initialization period of at least $10000 N$ periods of the dynamics for the system state to settle down to an attractor. In the case of type II chimera states, we observe extremely long oscillatory transients, during which the number of synchronized oscillators increases with time. The duration of the transients is found to scale exponentially with system size and is of the order of $10^6$ periods of the dynamics for $N=256$. For a give system size, we find that the time between successive oscillators joining the synchronized cluster grows exponentially with increasing size of the cluster. The observed dynamics might be related to super-transients as reported in~\cite{Wolfrum2011}, but a more detailed analysis of the transients is beyond the scope of this paper. In order to minimize the influence of transient dynamics on our studies, we chose the initialization period so long as to observe no more changes in the number of synchronized oscillators for at least $10000 N$ periods. Based on an analysis of Fourier spectra, we assume average period lengths of the dynamics of approximately $4.2$ time units for type I and $12.9$ for type II chimera states, independent of the system size $N$. For the computation of LEs and BOLVs, we employ the standard method by Eckman et al.~\cite{Eckmann1985}, which is based on repeatedly performing $QR$-decompositions and averaging over the logarithmic diagonal entries of the $R$ matrices for a sufficiently long time. We choose to perform 20 $QR$-decompositions per period of the dynamics. Expecting an exponential convergence of the $Q$-matrices containing the BOLVs, we admit a transient period of at least $1000 N$ periods before recording the LEs and BOLVs. The CLVs are obtained from the BOLVs and the respective $R$ matrices by applying the dynamic algorithm by Ginelli et al.~\cite{Ginelli2007}. In backward-time direction, we admit an initialization period of at least $100 N$ periods before recording the CLVs. The final estimate of the LEs is obtained from an average over at least $2.5\cdot10^4$ periods of the dynamics. The CLVs are saved after each $QR$-step (resulting in 20 samples per period) over a time interval of $2.5\cdot 10^4$ periods. Based on the statistics of the short-time estimates, we expect an accuracy of the exponents of at least $10^{-3}$.

\section{Results and Discussion\label{sec:results}}
\subsection{Lyapunov spectra for $N=16$ oscillators\label{sec:spectra}}
\begin{figure*}
	\includegraphics[]{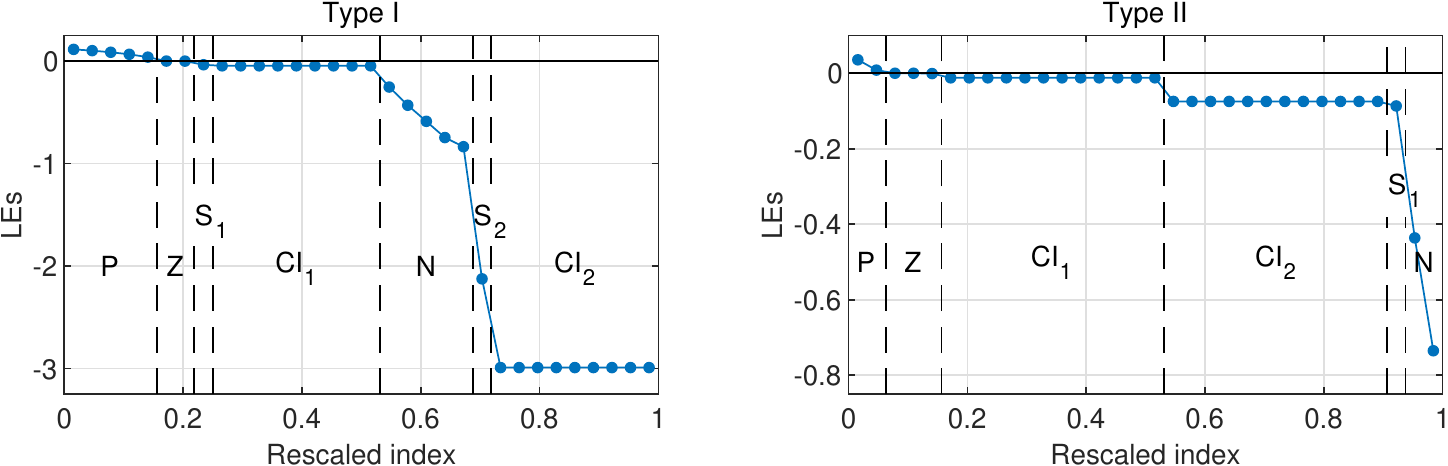}
	\caption{Lyapunov spectra of a type I chimera state in Eq.~\eqref{eq:slelgc} and a type II chimera state in Eq.~\eqref{eq:slenlgc} for a system size of $N=16$ oscillators.\label{fig:spectra:single}}
\end{figure*}

Fig.~\ref{fig:spectra:single} shows Lyapunov spectra of the type I and type II chimera states for a system size of $N=16$ oscillators. Simplifying further discussions, we plot the full spectra,  consisting of $2N = 32$ exponents, without removing degenerate exponents. We furthermore follow the suggestion of~\cite{Pikovsky2016} and introduce the rescaled index 
\begin{equation*}
	\tilde{l} := (l-1/2)/(2N)\in\left(0,1\right)\text{,}
\end{equation*}
which enables us to show spectra for different ensemble sizes over the same abscissa range. The simulation data, used to generate the spectra, is not depicted here, but resembles the data in Fig.~\ref{fig:dynamics} closely. For the type I dynamics, we find a group of $N_\text{sync} = 10$ synchronized oscillators, which coexists with $N_\text{inc} = 6$ oscillators performing incoherent motion. For the type II dynamics, the fraction of synchronized oscillators is higher with $N_\text{sync} = 13$ and $N_\text{inc} = 3$. The clustering of the oscillators into synchronized and incoherent groups induces a similar grouping pattern also for the LEs: These can be assigned to four main groups -- P, N and $\text{CI}_{1\text{,}2}$ --, which are separated by a small number of singleton exponents -- Z and $\text{S}_{1\text{,}2}$. 

Starting with P, the group of positive exponents, we recognize that both types of chimera possess $N_\text{inc}-1$ positive exponent, revealing their hyperchaotic character. In addition to P, we find a group N of negative exponents, with the same number of exponents, cf.~Fig.~\ref{fig:spectra:single}. The facts that P and N are identical in size, and that the number of exponents appears to relate to the number of incoherent oscillators in the ensemble, suggests that P- and N-exponents correspond to stable and unstable directions in tangent space, which affect the group of incoherent oscillators predominantly. The shapes of the corresponding CLVs support this finding. In particular, all CLVs associated with the P- and N-exponents perturb all synchronized oscillators in the same way, without disturbing the clustering pattern. 

Besides groups P and N, we find a group Z of very small exponents with values of the order of $10^{-4}$ or smaller. We observe two such exponents for the type I chimera state and three for the type II dynamics. These numbers are consistent with the number of zero-exponents to be expected from symmetry considerations of Eqs.~\eqref{eq:slelgc} and~\eqref{eq:slenlgc}. The governing laws of both systems are independent of time, so that for any solution $\wref\!\left(t\right)$ of Eqs.~\eqref{eq:slelgc} or~\eqref{eq:slenlgc} the time-shifted solution $\wref\!\left(t+\delta t\right)$ provides a valid trajectory, as well. Perturbations along the dynamical flow therefore neither grow nor decay in time, resulting in a neutrally stable direction of perturbation and a zero LE. Trivially, the invariance is preserved also when changing to the real-valued coordinates. With Eq.~\eqref{eq:ode} it is easy to see that the corresponding CLV is 
\begin{equation*}
	\delta \bm{x}_\text{ts} \propto \bm{f}\!\left(\xref\!\left(t\right)\right)\text{,}
\end{equation*}
which, by chain rule, satisfies Eq.~\eqref{eq:tanspace}. Similarly, Eqs.~\eqref{eq:slelgc} and~\eqref{eq:slenlgc} are invariant with respect to phase shifts in the complex plane, i.~e.\ for any angle $\phi\in\R$, $e^{i\phi}\,\wref\!\left(t\right)$ is a valid solution if the same is true for $\wref\!\left(t\right)$. As a result, another zero-exponent is associated with the covariant Lyapunov vector
\begin{equation*}
	\delta\bm{x}_\text{ps} \propto \left(-b_1,\ldots,-b_N, a_1,\ldots,a_N\right)^T\text{,}
\end{equation*}
corresponding to an infinitesimal rotation of all oscillators in the complex plane. 

For the type II chimera state in Eq.~\eqref{eq:slenlgc}, we have an additional zero exponent, which arises from the conservation law
\begin{equation}
		d/dt\left\lvert\left\langle W \right\rangle\right\rvert = 0\text{.}
		\label{eq:conslaw}
\end{equation}
This can be seen from the fact that
\begin{equation}
	d \left\langle W \right\rangle/dt = -\nu \left\langle W \right\rangle\text{,}
	\label{eq:harmonicmean}
\end{equation}
wherein $\left\langle W \right\rangle = \sum_{k=1}^{N} W_k/N$ denotes the complex amplitude of the mean field. According to Eq.~\eqref{eq:harmonicmean}, $\left\langle W\right\rangle$ performs a harmonic oscillation in time, and thus conserves the mean-field amplitude. The conservation law~\eqref{eq:conslaw} induces a splitting of phase space into invariant manifolds associated with different values of $\left\lvert\left\langle W \right\rangle\right\rvert = \eta$, and thus yields another neutrally stable direction in tangent space. The CLV associated with this direction, however, appears to possess a non-trivial structure so that an analytical expression has not been identified, yet.

Two further groups of exponents, $\text{CI}_{1,2}$, are made up of only two distinct LEs, possessing a $\left(N_\text{sync}-1\right)$-fold degeneracy, each. Clearly, these degeneracies originate from the synchronization pattern of the chimera states. To see this, we have to take a closer look on the tangent space dynamics in presence of synchronized oscillators. Extending the notation of Eq.~\eqref{eq:ode}, we write
\begin{equation*}
	\dot{\bm{x}} = \left(
	\begin{array}{c}
		\dot{\bm{a}} \\ 
		\dot{\bm{b}}
	\end{array}\right) 
	= \left(
	\begin{array}{c}
		\bm{f}^{(a)}\!\left(\bm{a}, \bm{b}\right) \\ 
		\bm{f}^{(b)}\!\left(\bm{a}, \bm{b}\right)\text{.}
	\end{array}\right) = \bm{f}\!\left(\bm{x}\right)\text{,}
\end{equation*}
with $\bm{a}=\left(a_1, \ldots, a_N\right)^T$ and $\bm{b}=\left(b_1,\ldots,b_N\right)^T$. The Jacobian matrix of the overall system can then be written in block-matrix form as
\begin{equation}
\renewcommand{\arraystretch}{2.5}
	\dfrac{\partial\bm{f}}{\partial\bm{x}} = \left(
	\begin{array}{c | c}
	\dfrac{\partial\bm{f}^{(a)}}{\partial\bm{a}} & \dfrac{\partial\bm{f}^{(a)}}{\partial\bm{b}} \\ \hline
			\dfrac{\partial\bm{f}^{(b)}}{\partial\bm{a}} & \dfrac{\partial\bm{f}^{(b)}}{\partial\bm{b}}
	\end{array}\right)\text{.}
	\label{eq:jacobian}
\end{equation} 
Denoting the $k^\text{th}$ coordinate of $\bm{f}^{(a)}$ with $f_k^{(a)}$, the global coupling scheme in Eqs.~\eqref{eq:slelgc} and~\eqref{eq:slenlgc} allows us to separate the terms according to
\begin{equation}
	f_k^{(a)}\!\left(\bm{a},\bm{b}\right) = g^{(a)}\!\left(a_k, b_k\right) + \sum_{j=1}^{N} c^{(a)}\!\left(a_j,b_j\right)\text{,}
	\label{eq:partition}
\end{equation}
with $c^{(a)}$ and $g^{(a)}$ being independent on the oscillator indices $k$ and $j$, respectively. A similar argument is valid for $\bm{f}^{(b)}$. Sorting then the oscillators in a way that synchronized oscillators $W_k$ obtain indices $k=1,\ldots,N_\text{sync}$, we can conclude from Eq.~\eqref{eq:partition} that each of the sub-Jacobians in Eq.~\eqref{eq:jacobian} can again be written in block-matrix form as
\begin{equation*}
\renewcommand{\arraystretch}{1.5}
	\frac{\partial \bm{f}^{(p)}}{\partial \bm{q}} = \left(
		\begin{array}{c | c}
		g_{pq} \bm{1} + c_{pq} E & \ \dots \\ \hline
		R_{pq} & \ \dots
		\end{array}\right)\text{,}
\end{equation*}
wherein for $p,q\in\left\{a,b\right\}$, 
\begin{align*}
	g_{pq} &:= \left.\frac{\partial g^{(p)}}{\partial q}\right\rvert_{a_1,b_1}\text{,} \\
	c_{pq} &:= \left.\frac{\partial c^{(p)}}{\partial q}\right\rvert_{a_1,b_1}\text{,}
\end{align*}
and $\bm{1},E\in\R^{N_\text{sync}\times N_\text{sync}}$ denote the unit matrix and the matrix with entries all equal to unity. Furthermore, the entries of $R_{pq}\in\R^{N_\text{inc}\times N_\text{sync}}$ are row-wise constant. From this structure, we can deduce that for perturbations affecting only the relative position of synchronized oscillators, i.~e.\
\begin{align}
\delta \bm{x}_\perp := \left(\right.&\delta a_1,\ldots,\delta a_{N_\text{sync}},0,\ldots,0,\ldots \nonumber \\
&\delta b_1,\ldots,\delta b_{N_\text{sync}},0,\ldots,0 \left.\right)^T\text{,}
\label{eq:syncpert}
\end{align}
with $\sum_{k=1}^{N_\text{sync}}\delta a_k = \sum_{k=1}^{N_\text{sync}}\delta b_k = 0$, we have
\begin{equation*}
\sum_{k=1}^{N_\text{sync}} d\left(\delta a_k\right)/dt = \sum_{k=1}^{N_\text{sync}}d \left(\delta b_k\right)/dt = 0\text{,}
\end{equation*}
as well as
\begin{equation*}
d\left(\delta a_k\right)/dt = d\left(\delta b_k\right)/dt = 0\text{,}
\end{equation*}
for $k = N_\text{sync}+1,\ldots,N$. Perturbations of type~\eqref{eq:syncpert} thus form an invariant subspace under time evolution in tangent space, the dimension of which is $2 N_\text{sync}-2$. Reducing tangent space dynamics to only this subspace, it is possible to obtain a two-dimensional system of linear equations, whose temporal asymptotics exactly reproduce the values of the two degenerated LEs, as obtained from the full system dynamics. Similar sets of exponents have also be found by Ku et al.~\cite{Ku2015} in the context of pure cluster states in oscillatory systems with linear global coupling, and were termed cluster integrity exponents. Due to the very similar role of the degenerate exponents in our context, we adopt this nomenclature here, leading to the group annotations $\text{CI}_{1}$ and $\text{CI}_{2}$.

What remains in both spectra is a small number of singleton exponents -- $\text{S}_1$ and $\text{S}_2$ in the type I spectrum and only $\text{S}_1$ for type II --, the role of which will become clear when investigating the localization properties of the associated covariant Lyapunov vectors.

\subsection{Lyapunov spectra for larger systems}

\begin{figure*}
	\includegraphics[]{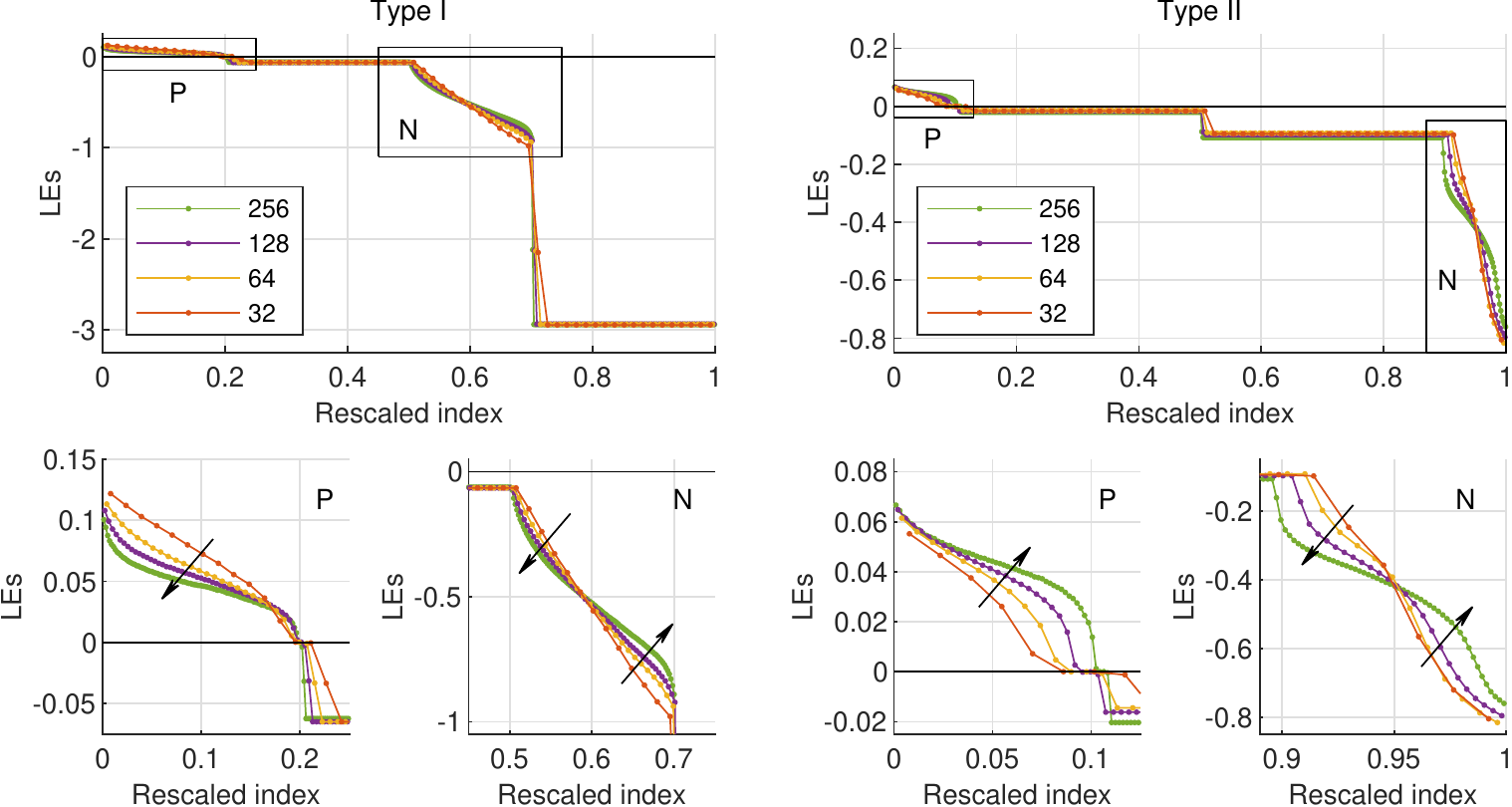}
	\caption{Lyapunov spectra of type I and type II chimera states for increasing ensemble sizes from $N=32$ to $N=256$.\label{fig:spectra:full}}
\end{figure*}

Fig.~\ref{fig:spectra:full} shows Lyapunov spectra for type I and type II chimera states in Eqs.~\eqref{eq:slelgc} and~\eqref{eq:slenlgc}. Again, the spectra are plotted against the rescaled index and include the degenerate exponents. Apart from small fluctuations of the order of two to four oscillators, the number $N_\text{inc}$ of incoherent oscillators scales roughly linearly with ensemble size, so that the structure of the spectra is similar to those in Fig.~\ref{fig:spectra:single}. In particular, we find that in most cases the number of P- and N-exponents equals $N_\text{inc}-1$ each, and thus scales extensively with system size. Only for the type I spectra with $N=128$ and $N=256$ oscillators, there are $N_\text{inc}-2$ positive exponents and one additional singleton exponent very close to zero. This, however, is likely to be an artifact. On the one hand, especially for high-dimensional systems, it may take a long time for the smallest exponents to converge properly, since these correspond naturally to the slowest directions of perturbation. On the other hand, the $QR$-based method for computing LEs is known to be vulnerable to numerical errors in computing close-to-zero exponents and associated BOLVs~\cite{Dieci1997,Dieci2005,Pikovsky2016}.

Similar to P- and N-exponents, the groups of degenerate exponents scale extensively with system size, as well. In agreement with theory, we find a degeneracy of degree $N_\text{sync}-1$ for each of the exponents.

While the values $\lambda_l$ of the degenerate exponents do not appear to follow an obvious trend as a function of system size, the P- and N-exponents exhibit a pronounced size dependence with a significant trend. To give a better overview on the size dependence of the non-degenerate P- and N-exponents, the smaller sub-figures in Fig.~\ref{fig:spectra:full} depict magnified views onto the relevant regions. For the type I dynamics, we observe a decreasing trend for the main part of the positive exponents while some of the positive exponents close to the zero group appear to follow an increasing trend. The bulk of P-group thus seems to approach a constant value, which is consistent with former results by Takeuchi et al.~\cite{Takeuchi2013}. However, for the most positive exponents, the rate of decrease is slower compared to bulk and furthermore appears to slow down with increasing $N$. Similar to the observations of Takeuchi et al., this might indicate a sub-partitioning of the positive exponents into an extensively scaling and a sub-extensive group. 

In contrast to the type I case, we observe an increasing trend for the main part of the positive exponents within the type II spectra. Only some of the smallest positive exponents do not follow the increase. In particular, the smallest positive exponent seems to approach zero, so that for $N=256$ oscillators four instead of three exponents are very close to zero. From a numerical perspective, the values of the Z-group exponents of the type II dynamics fluctuate stronger compared to the those of the type I spectra. This might indicate difficulties with the numerical integration method. Still, it might also give an indication that in the limit of large system sizes some of the CLVs come very close to those of the zero subspace and should thus be explored further. 

The N-group of negative exponents appears to flatten with increasing system size for both types of chimera states, i.~e.\ more negative exponents increase whereas less negative exponents decrease. For the type I spectra, this results in an increasingly tangent-like shape of the spectral curve. A similar pattern is found also within the type II spectra. A difference, however, occurs in the behavior of the most negative exponents between $\tilde{l}=0.95$ and $\tilde{l}=1$. Within this range, the type II spectra deviate significantly from the tangent-like shape. The presence of an additional turning point in the spectral curve around $\tilde{l}=0.975$ results in an step-like shape of the N-group spectrum and might suggest that there exist additional substructures in the spectrum, which are not resolved by the grouping structure proposed in this paper.

\subsection{Lyapunov dimension and extensivity of type I and type II chimera states} 
  
 \begin{figure}
	\includegraphics[]{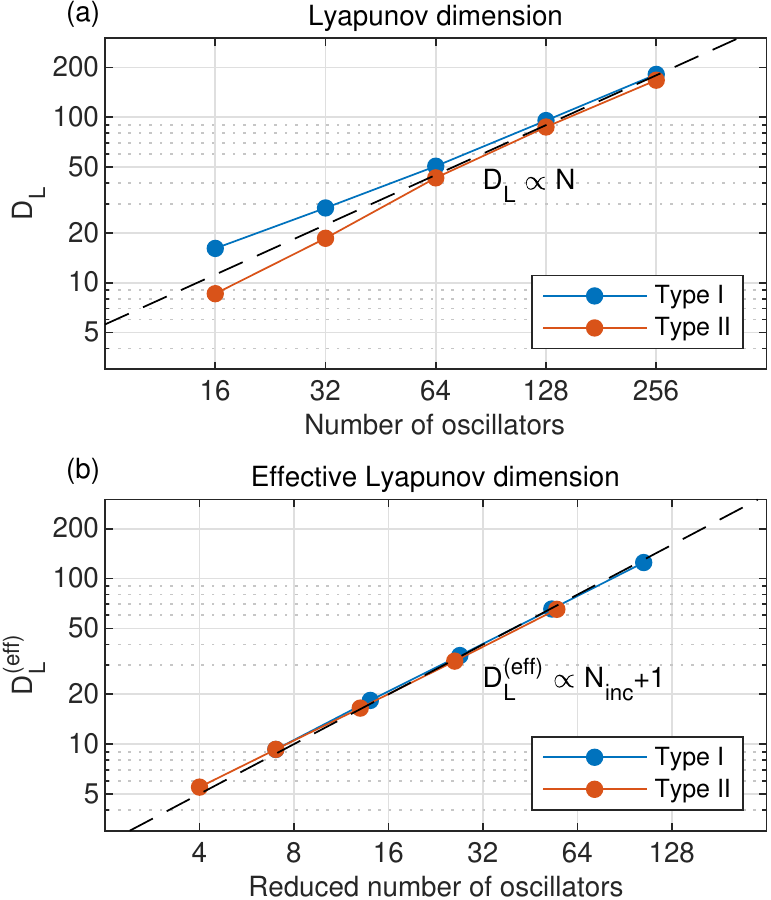}
	\caption{(a) Lyapunov dimensions type I and a type II chimera states as a function of $N$. (b) Effective Lyapunov dimension as a function of the reduced number of oscillators $N_\text{inc}+1$.\label{fig:dimensions}}
\end{figure}

Given the Lyapunov spectra, it is straight forward to compute also the Lyapunov dimension as stated in Eq.~\eqref{eq:lyapdim}. The computed values are shown in Fig.~\ref{fig:dimensions} (a) and reveal a monotonic increase of $D_\text{L}$ with system size. Interestingly, the trend of the type I chimera state (at least for system sizes below $N=256$) is not perfectly linear, but slightly slower. Going to larger system sizes, however, the deviation from linear growth decreases, suggesting that the attractor dimension of the chimera state behaves extensively in the limit $N\to\infty$. Similarly, the Lyapunov dimension appears to grow slightly faster than linear for type II chimera states, but growth slows down when considering larger systems.

The extensive nature of the attractor dimension is further strengthened by another consideration. Due to identical synchronization of $N_\text{sync} = N-N_\text{inc}$ oscillators, the system state $\xref\!\left(t\right)$ is necessarily confined to a $2(N_\text{inc}+1)$-dimensional subspace of the $2 N$-dimensional phase space. From Eqs.~\eqref{eq:slelgc} and~\eqref{eq:slenlgc}, it is easy to see that oscillators remain synchronized for all time, once they move in synchrony, so that this subspace is invariant under the dynamical flow. Although simulations achieve synchronization only within numerical accuracy, it thus seems natural to compute the Lyapunov dimension in a way that excludes directions orthogonal to the synchronization manifold. In section~\ref{sec:spectra}, we have seen that these directions correspond to the cluster integrity exponents $\text{CI}_1$ and $\text{CI}_2$. Therefore, removing the $2(N_\text{sync}-1)$ degenerate exponents before computing $D_\text{L}$ results in an effective value of the Lyapunov dimension, $D_\text{L}^\text{(eff)}$, which respects the synchronization pattern. This procedure is equivalent to considering an $(N_\text{inc}+1)$-oscillator system, in which the synchronized oscillators are replaced by a single representative and weighted accordingly before taking averages. When plotting this effective dimension as a function of the reduced system size $(N_\text{inc}+1)$, finite-size effects are minimized and the deviations from linear growth are marginal, as shown in Fig.~\ref{fig:dimensions}~(b). To quantify the deviation, we presumed a power-law dependence of the form
\begin{equation*}
	D_\text{L}^\text{(eff)} = \alpha (N_\text{inc}+1)^\beta\text{,} 
\end{equation*}
and obtained power-law exponents $\beta=0.96$ for the type I chimera and $\beta=0.94$ for the type II chimera from a log-log linear fit to the data. Within numerical accuracy, both exponents are well consistent with an extensive scaling of the effective Lyapunov dimension, and thus with extensively chaotic motion of the incoherent oscillators. 

\subsection{Localization of covariant Lyapunov vectors\label{sec:localization}}

\begin{figure*}
	\includegraphics[]{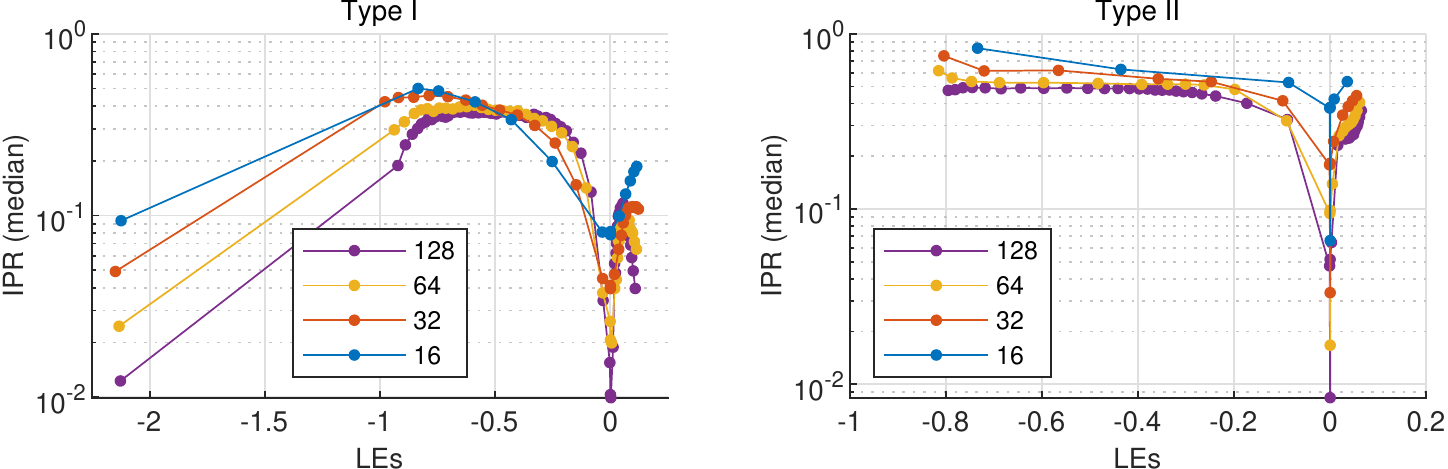}
	\caption{Median localization of covariant Lyapunov vectors.\label{fig:localization}}
\end{figure*}

As a last part of the Lyapunov analysis of type I and type II chimera states, we discuss the localization properties of the CLVs in terms of the IPR as stated in Eq.~\eqref{eq:ipr}. Generically, due to time dependence of the CLVs themselves, we find that the IPR value for a given vector fluctuates strongly as a function of time. This fact has already been noted by Takeuchi et al.~\cite{Takeuchi2009}, who considered the temporal mean value $\left\langle \text{IPR}\right\rangle_t$ in order to overcome this problem. Within our study, we changed this approach slightly and replace the mean with the median of the temporal distribution. Investigating the distribution of IPR values in time, we noticed that the mode of the probability distribution especially for larger ensemble sizes and delocalized modes is typically situated at values that are notably smaller than the temporal mean of the distribution. Increasing the ensemble size, we found that the separation between mode and mean grows monotonically, so that vector delocalization appears to be underestimated systematically by the mean value for larger system sizes. As a reason for this effect, we identified an extremely heavy-tailed shape of the distribution functions, which favors larger values of IPR in a way that introduces bias to the mean. Employing the median of the temporal distribution instead of the mean reduces the dependence on heavy tails and is expected to yield more reliable results. In Fig.~\ref{fig:localization}, we therefore show the temporal median IPR of the CLVs, plotted as a function of the corresponding LEs. Therein, special care had to be taken for the cluster integrity exponents $\text{CI}_1$ and $\text{CI}_\text{2}$, as well as for the zero-exponents Z. Since all of the CLVs associated with a degenerate exponent share the same invariant subspace in tangent space, and due to linearity of the tangent space dynamics, any linear combination of two or more CLVs is again a valid CLV with the same exponent. The IPR, however, is intrinsically nonlinear and varies strongly when performing linear combinations, so that strictly speaking the IPR can not be properly defined in degenerate subspaces. In practice, of course, well-defined values are obtained, but the values depend on initialization details and thus carry no information on the actual dynamics. For reasons of clarity, we thus excluded the groups $\text{CI}_1$ and $\text{CI}_2$ from the plots. For the zero-exponents, especially for the type II chimera state, we observed a similar effect. Knowing analytical solutions of the corresponding covariant Lyapunov vectors, however, we chose one vector to be equal to the time derivative of the dynamics $\delta\bm{x}_\text{ts}$, another one to equal the phase-shift direction $\delta\bm{x}_\text{ps}$, and the remaining one to lie within the zero-subspace, linearly independent of the former ones. 

From the localization spectra, we recognize that the grouping of the Lyapunov spectra manifests itself also in the localization properties of the CLVs. Regarding the type I spectra in Fig.~\ref{fig:localization}, we observe a notable decrease of the IPR with increasing system size for the largest few exponents, for close-to-zero exponents and for some of the smallest exponents. These findings are consistent with the results of Takeuchi et al.~\cite{Takeuchi2009,Takeuchi2013}, who identified five collective Lyapunov modes in similar positions for string-like chaotic states in the same set of equations. Trivially, the zero-exponents, resulting from time- and phase-shift symmetries, are associated with delocalized vectors, yielding two neutrally stable collective Lyapunov modes. Close to zero, however, we observe additional Lyapunov modes that appear to become increasingly delocalized with increasing system size and correspond to both positive and negative near-zero exponents. On the negative side, we observed already in Fig.~\ref{fig:spectra:single} the close-to-zero singleton exponent $\text{S}_1$, which possesses analogues also in spectra for larger system sizes, as visible in Fig.~\ref{fig:spectra:full} (Type I, P). The exceptional position of this exponent in the spectra might indicate correspondence to some collective Lyapunov mode or to an otherwise special direction in phase space yet to be identified. Similarly, we observe also a small number of positive exponents, for which the median IPR appears to decrease with increasing $N$. However, these close-to-zero modes are hard to separate from the zero-subspace because of both, strong finite-size effects, which were also found by Takeuchi et al.~\cite{Takeuchi2009}, and the before-mentioned inherent numerical inaccuracies of close-to-zero exponents and associated BOLVs from $QR$-decomposition. We thus leave the study of larger systems, which exceed our current computational facilities, for future research. Nevertheless, the qualitatively similar pattern of Lyapunov spectra and median respectively mean IPRs of the incoherent group of type I chimera states and string-like chaos in ensembles of SL oscillators with linear global coupling reveals that the synchronized group plays only an inferior role for the macroscopic chaotic motion of the ensemble. This could be a hint that the fast collective modes (corresponding to the most stable and unstable CLVs) are only delocalized among the incoherent oscillators while the synchronized group is not affected. 
Looking at the spectrum of the type II chimera state, it appears flatter than the one for the type I dynamics, and a notable decrease of the median IPR is visible only for zero exponents and the ones close-by, cf.~Fig.~\ref{fig:localization}~(Type II). In particular, we note that in the type II spectrum neither the most positive LE nor the singleton exponent $\text{S}_1$ appear to be associated with a collective Lyapunov mode, since their median IPR decrease only slightly. This reveals significant differences between the collective properties of type I and type II chimera states. It is tempting to conclude that it is the additional constraint on the dynamics of type II chimeras, the conservation of the mean oscillation~\eqref{eq:conslaw}, that impedes the formation of fast collective motion. However, at the current state of analysis this remains a hypothesis. A starting point for further investigations are type-I-like chimera states which transiently form also in Stuart-Landau ensembles subject to nonlinear global coupling \cite{Schmidt2015}. If for some parameter values these states can be stabilized or at least pushed to a 'super-transient regime', a Lyapunov analysis should reveal whether also then the collective modes are suppressed, or whether the string-like chaos of the incoherent group is a signature of its collective dynamics.

\section{Conclusion\label{sec:discussion}}

In the present paper, we have investigated Lyapunov spectra and localization properties of the associated covariant Lyapunov vectors (CLVs) for two examples of chimera states in systems of globally-coupled amplitude oscillators. We found that for both type I and type II chimera states the Lyapunov exponents can be subdivided into four main clusters, containing extensively many LEs, and a small number of singleton and zero exponents. Two of the main clusters, $\text{CI}_1$ and $\text{CI}_2$, are found to consist of degenerate exponents, which are a result of the synchronization pattern of the chimera state. The corresponding perturbations affect the synchronized oscillators only, leading to the name cluster integrity exponents. A further group contains the positive exponents and indicates the hyperchaotic nature of the chimera states. For both type I and type II chimeras, the spectra exhibit a pronounced dependence on system size. While, for the type I dynamics, most of the positive LEs decrease with increasing system size towards a limiting value, the main part of the positive exponents grows to a limiting value in the case of type II dynamics. The origin of this difference in behavior has to be addressed in future research. 

Notwithstanding the different trend of the positive LEs of our two types of chimera states, the Lyapunov dimension of both dynamics grows extensively with system size. Next to the standard Lyapunov dimension, computed from the whole spectrum of LEs, we defined an effective Lyapunov dimension by neglecting the cluster integrity exponents, which measures the dimensionality only of the effective dynamics and takes account for the synchronization pattern. The size dependence of the effective Lyapunov dimension can be fitted with a power law, and yields a power-law exponent very close to unity for both type I and type II dynamics.

An analysis of the localization properties of covariant Lyapunov vectors for type I and type II chimera states revealed further differences between the dynamics. While we found evidence for at least four collective Lyapunov modes being present in the type I dynamics, two of them being associated with large positive and negative LEs, the data suggests that no more than four weak collective modes with LEs zero or close to zero exist in the type II chimera state. Future research should set in at this point and provide additional data for larger system sizes to exclude finite size effects from the localization spectra and to determine the exact number of collective Lyapunov modes for both types of chimera state. Here, it would be of particular interest whether the collective modes form a sub-extensive group that grows approximately as $\mathcal{O}\!\left(\ln\!\left(N\right)\right)$ as found for string-like chaos~\cite{Takeuchi2013}.

From a theoretical point of view, it can be expected that constraints on the system dynamics have a qualitative impact on tangent-space geometry. It therefore appears promising to investigate the geometric relation between the collective Lypaunov modes, for example in terms of angles or correlations between the vectors. Such a study could reveal information on the effective degrees of freedom within the ensemble, and might thus guide a way towards a statistical description of nonlinear amplitude oscillators subject to global coupling and constraints. To address this topic, it furthermore appears reasonable to examine in which way the collective modes affect the oscillator distribution in the complex plane. Solutions to these problems, however, have to be based on a thorough statistical analysis of the dynamics and thus require the study of oscillatory ensembles of much larger size.

+\section*{Acknowledgments}

The authors thank S. W. Haugland for fruitful discussions.
The project was funded by the Deutsche Forschungsgemeinschaft (DFG, German Research Foundation) project KR1189/18.
\bibliography{lit}{}
\end{document}